\documentclass[pra
,twocolumn
]{revtex4-1}

\usepackage{graphicx}
\usepackage{calc}
\usepackage{braket}
\usepackage{amsmath}
\usepackage{amssymb}

\newcommand{\ep}{\epsilon^\prime}
\begin{document}

\title{Generation of distributed entangled coherent states over a lossy
  environment with inefficient detectors}

\author{A.~P.~Lund}
\author{T.~C.~Ralph}
\affiliation{Centre for Quantum Computation and Communication Technology,\\
The University of Queensland}

\author{H.~Jeong} 
\affiliation{Center for Macroscopic Quantum Control and\\
  Department of Physics and Astronomy,\\
  Seoul National University, Seoul 151-747, Korea}

\begin{abstract}
  Entangled coherent states are useful for various applications in quantum
  information processing but they are are sensitive to loss.  We propose a
  scheme to generate distributed entangled coherent states over a lossy
  environment in such a way that   the fidelity is independent of the losses at
  detectors heralding the generation of the entanglement.  We compare our
  scheme with a previous one for the same purpose [Ourjoumtsev {\em et al.},
  Nat. Phys. {\bf 5} 189 (2009)] and find parameters for which our new
  scheme results in superior performance.
\end{abstract}

\maketitle

\section{Introduction}

Entangled coherent states (ECSs) \cite{MecozziTombesi,Sanders92,SandersReview}
are useful for various applications in quantum information processing
\cite{vanEnkPRA2001,JeongPRA2001,Gerry2001,Wang2001,JeongPRA2002,JeongQIC2002,
	RalphPRA2003,An2003,JeongAn2006,LundPRL2008,AnKim2009,Munhoz2010,Joo2011,MyersRaph2011,
Joo2012,Kirby2013,Jonas2013} and for exploring fundamental properties of quantum theory
\cite{GerryGrobe1995,GerryGrobe1997,Wilson2002,JeongSon2003,Munro2000,
Stobinska2007,JeongPRL2009,Paternostro2010,Lee2011,Tor2013}.  An optical ECS can be
generated by passing a single-mode superposition of coherent states (SCS)
\cite{Milburn1985,YurkeStoler,Schleich} through a beam splitter.  SCSs in
free-traveling fields have been experimentally generated in several
laboratories
\cite{Ourjoumtsev2006,OurjoumtsevNature2007,SasakiPRL2008,Neer2010,Gerrits2010}.  In order
to use an ECS for quantum communication or faithful nonlocality tests, it
should be distributed to two parties that are spatially separated.  However,
ECSs are sensitive to a lossy environment, a property typical of macroscopic
entanglement.  A method based on quantum repeaters for ECSs~\cite{ECSrepeater}
may not be efficient due to the requirement of efficient
photon-number-resolving detectors.  It would thus be very useful to develop a
scheme to efficiently generate an ECS distributed to two separate parties over
a lossy environment.

Ourjoumtsev {\it et al.}  experimentally demonstrated such a method for
generating a distributed ECS~\cite{grangier:nature}. They utilized the fact
that if an ECS is in an asymmetric form with different amplitudes for the two
modes, the mode with the smaller amplitude is relatively more robust against
loss.  However, their approach~\cite{grangier:nature} is sensitive to
inefficiency of the photon detector used for the necessary post-selection, and
this formed a major part of the limitations in its practical applications.

In this paper, we present a scheme to generate distributed ECSs over a lossy
environment whose output fidelities with ideal ECSs are insensitive to
detection inefficiency.  Our scheme employs the loss tolerant construction of
Refs.~\cite{JeongPRA2002,lund:catboost} to achieve the robustness against lossy
detection.  We compare output fidelities obtained by our scheme with those of
the previous one~\cite{grangier:nature}.  We also show that there exists a
range of parameters for which our scheme provides better results even when
taking into account the lower success probability of the new scheme.

This paper is organized as follows. In Sec.~\ref{sec:unam}, we briefly review
the loss tolerant detection for coherent state qubits~\cite{JeongPRA2002},
which is employed as an important part in our scheme.  In
Sec.~\ref{sec:grangier}, we review and analyze the scheme in
Ref.~\cite{grangier:nature}.  Our scheme is then presented in Sec.~III with a
comparison between the two proposals in terms of the success probabilities and
the fidelities.  We conclude our paper with final remarks in Sec.~IV.

\section{Unambiguous and loss tolerant detection for coherent states}
\label{sec:unam}
 
\begin{figure}[hb!]
  \includegraphics[width=1.8in]{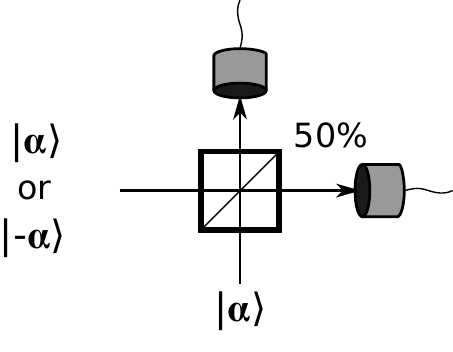}
  \caption{Loss tolerant detection scheme from Ref.~\cite{JeongPRA2002}.  A
  coherent state qubit encoded in the basis $\ket{\alpha}$ and $\ket{-\alpha}$
  is interfered with a coherent state of the same amplitude at a 50:50 beam
  splitter.  Perfect classical interference allows for the two input
  coherent states to be distinguished without error by recording which detector
  registers a click.  However, due to the overlap of the two possible inputs,
  this measurement must be heralded and have some non-zero probability of
  failure.  Failure occurs when no clicks are recorded.  Detector loss
  increases this failure rate but does not induce errors in discrimination
  between the two basis states when photons are detected.}
  \label{detection_scheme}
\end{figure}

As a strategic part of our construction, we will first review unambiguous and
loss tolerant detection for coherent state qubits discussed in
Ref.~\cite{JeongPRA2002}.  This method was used for purifying
ECSs~\cite{JeongQIC2002} and amplifying SCSs~\cite{lund:catboost}.  Suppose
that one needs to unambiguously discriminate between two coherent states,
$\ket{\pm\alpha}$ with amplitudes $\pm\alpha$, using two inefficient photon
detectors.  A simple way to perform this task is to apply a 50:50 beam splitter
to interact the coherent state with an ancillary coherent state $\ket{\alpha}$
as shown in Fig.~\ref{detection_scheme}.  If the input system mode was in state
$|\alpha\rangle$, we get the transformation
\begin{equation}
  \ket{\alpha,\alpha} \rightarrow \ket{\sqrt{2}\alpha,0},
\end{equation}
and if the system was in state $\ket{-\alpha}$, the transformation is
\begin{equation}
  \ket{-\alpha,\alpha} \rightarrow \ket{0,\sqrt{2}\alpha}.
\end{equation}
This is an example of perfect classical interference and consequently only one
of the two output modes contains any photons.  It is then clear that only
one of the two detectors in Fig.~\ref{detection_scheme} can click for the
final detection and such a click will unambiguously identify the initial state
of the system.  Of course, the signal we are trying to detect
(i.e. $\ket{\pm\sqrt{2}\alpha}$) has a non-zero overlap with the vacuum
state.  This corresponds to the failure probability in which both the detectors
are silent.  It is important to note that inefficiency of the detectors will
only increase this failure probability but it does not affect the unambiguous
nature of the detection scheme.  If we can ignore other experimental issues
such as mode-mismatching at the beam splitter and dark counts at the detectors,
the result will always be reliable as far as one of the detectors registers any
photon(s).

\section{Entanglement distribution using photon number detection}
\label{sec:grangier}

Writing the Bell basis in the form of ECSs gives
\begin{equation}
N_{\pm}(\sqrt{2}\alpha) \left( \ket{\alpha,\alpha} \pm 
\ket{-\alpha,-\alpha} \right),
\end{equation}
where the other two Bell states are generated by applying a local $\pi$ phase
shift to one mode.  The task considered in this paper is generating a state of
this form between two spatially separated parties that share a quantum channel,
can perform local measurements and can communicate the results of these
measurements via a classical channel.

\begin{figure}[h!]
  \centerline{\includegraphics[width=3in]{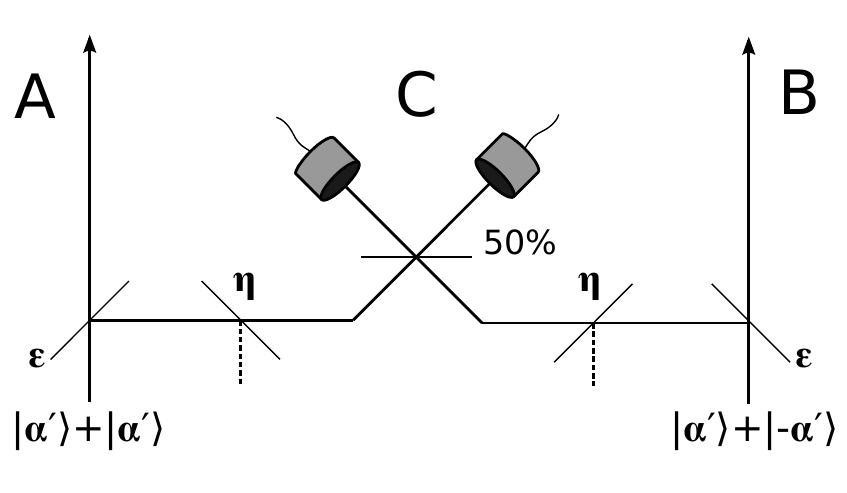}}
  \caption{Entanglement distribution scheme from Ref.~\cite{grangier:nature}.  A
    and B are the separated parties generating an ECS and C is a central
    herald.  The parameters shown here are $\eta$ the one sided loss,
    $\epsilon$ the tapping off ratio and $\alpha^\prime$ defined in
    Eq.~(\ref{aprime}).  These parameters are set so that the entanglement
    generated is a coherent state encoded Bell state with amplitude $\alpha$.
    Each detector has a loss of $l$ (not shown).}
  \label{grangier_experiment}
\end{figure}

Reference~\cite{grangier:nature} proposes and implements a method for
generating distributed ECSs.  The schematic for this scheme is shown in
Fig.~\ref{grangier_experiment}.  The scheme requires two parties, here called
A and B, to initially generate SCSs both with an encoding amplitude
$\alpha^\prime$.  Then both A and B use a beam-splitter to distribute a small
fraction of energy $\epsilon$ from their SCS and send it to a central
location C.  To ensure that the entanglement generated has a particular
coherent state encoding amplitude $\alpha$, it is necessary to set
\begin{equation}
	\label{aprime}
	\alpha^\prime = \sqrt{1+\epsilon^\prime} \alpha	
\end{equation}
where
\begin{equation}
	\epsilon^\prime = \frac{\epsilon}{1+\epsilon}.
\end{equation}
A loss of $\eta$ is present between A and C as well as B and C and it is an
assumption of this analysis that losses are distributed evenly.  This results
in a total loss between A and B of
\begin{equation}
	\label{etat}
	\eta_T = 1-(1-\eta)^2.
\end{equation}
At C the two low amplitude states are received and then interfered on a $50:50$
beam-splitter.  Entanglement generation between A and B is heralded by the
detection of a click at one of the outputs from the beam-splitter at C.  Local
corrections are applied by A and B depending on information C broadcasts about
which mode a detection has been registered in and the detection outcome parity.

The probability of success and the fidelity with the ideal entangled coherent
state can be computed and details of this calculation are shown in
Appendix~\ref{appendix:original}.  To simplify the expressions a little we
write
\begin{equation}
\eta^\prime = \sqrt{\eta} + l - l\sqrt{\eta}
\end{equation}
where $l$ is the detector loss.  Having these parameters combine like this is
expected as the scheme treats detector loss and channel loss similarly due to
the symmetry of the losses and the linear beam-splitter interaction.  We split
the probability and fidelity expressions into the two cases of the herald
detecting an even and odd number of photons in one detector.   The expressions
for fidelity and probability of generation are
\begin{widetext}
\begin{equation}
	\mathcal{F}_{even} = 
	(1+\tanh(2\ep\eta^\prime\alpha^2)\tanh(2\alpha^2))^{-1}
	\label{eq:fe}
\end{equation}
\begin{equation}
        P_{even} = \frac{e^{-2 \ep \alpha^2}}{(1+e^{-2 (1+\ep) \alpha^2})}
        \left( \cosh(2 \ep (1 - \eta^\prime)\alpha^2) - 1 \right)
        \left(
        \cosh(2 \ep \eta^\prime \alpha^2) (1+e^{-2\alpha^2}) +
        \sinh(2 \ep \eta^\prime \alpha^2) (1-e^{-2\alpha^2})
        \right)
       	\label{eq:pe}
\end{equation}
\begin{equation}
	\mathcal{F}_{odd} = 
	(1+\tanh(2\ep\eta^\prime\alpha^2)\coth(2\alpha^2))^{-1}
	\label{eq:fo}
\end{equation}
\begin{equation}
        P_{odd} = \frac{e^{-2 \ep \alpha^2}}{(1+e^{-2 (1+\ep) \alpha^2})}
        \sinh(2 \ep (1 - \eta^\prime)\alpha^2) 
        \left(
        \sinh(2 \ep \eta^\prime \alpha^2) (1+e^{-2\alpha^2}) +
        \cosh(2 \ep \eta^\prime \alpha^2) (1-e^{-2\alpha^2})
        \right)
	\label{eq:po}
\end{equation}
\end{widetext}
where for each probability we have summed over the two possibilities of which
detector measures the photons and the output state is assumed to have had any
local phase shift correction applied. 

The key idea for this scheme which builds resilience to channel loss is that
the energy distributed is small and hence the chance of loosing a quanta of
energy is also small.  For each quanta of energy lost a sign flip ($Z$) error
is introduced in the coherent-state basis, therefore if $\epsilon$ is small the error rate can be low.
This can be seen in the fidelity equations by allowing $\epsilon \ll \alpha^2$
and $\epsilon \ll l^\prime \alpha^2$.  However, under these conditions, the
probability of getting the heralding signal is also small.  These losses can be
a roadblock to implementing scaled-up protocols~\cite{ECSrepeater}.

\section{Proposed scheme with realistic on/off detectors}

We will now construct a new scheme which uses the loss tolerant unambiguous
state discrimination as a herald and maintains the feature of distributing a
very low amplitude through the lossy channel.

The two parties A and B, generate SCSs as before but of different
amplitude as this new protocol is asymmetric between the two parties. Here A is
the sender.  Party A generates a cat-state state of amplitude
$\sqrt{1+\epsilon^\prime} \alpha$ and sends a proportion $\epsilon$ along the
lossy channel $\eta_T$ to the receiving party B.  Party B then combines the
small received amplitude with a cat-state state of amplitude
\begin{equation}
	\label{app}
	\alpha^{\prime \prime} = \sqrt{\rho}\alpha
\end{equation}
on a beam-splitter with reflectivity $\rho$ where
\begin{equation}
	\label{rho}
	\rho=\epsilon^\prime(1-\eta_T).
\end{equation}
The heralding is performed by B who takes one of the output modes of the
beam-splitter mixes that with a coherent state of amplitude
$\sqrt{\gamma}\alpha$ where
\begin{equation}
	\label{gammaeqn}
	\gamma = 4 \rho(1-\rho).
\end{equation}
All of these parameters are chosen so that the final output that A and B will
share is an ECS with encoding amplitude $\alpha$ and so that for the states
where the coherent state phases A and B share are the same, the state input
into the detector at B is the vacuum state.  There are only three free
variables which determine all parameters at this point, $\alpha$, $\epsilon$
and $\eta_T$.  Successful entanglement generation is heralded when a detection
occurs (of any non-zero number of photons) in one of the two outputs from this
final beam-splitter with detector losses of $l$.  Note that there is now no
central party which acts as a herald as this is done by the receiver B who must
communicate to A which events were successful.  The successful output state is
an ECS with a positive superposition and in phase coherent states.  This is the
same state produced with the previous scheme on achieving an even parity
success result.
\begin{figure}[ht]
  \centerline{\includegraphics[width=3in]{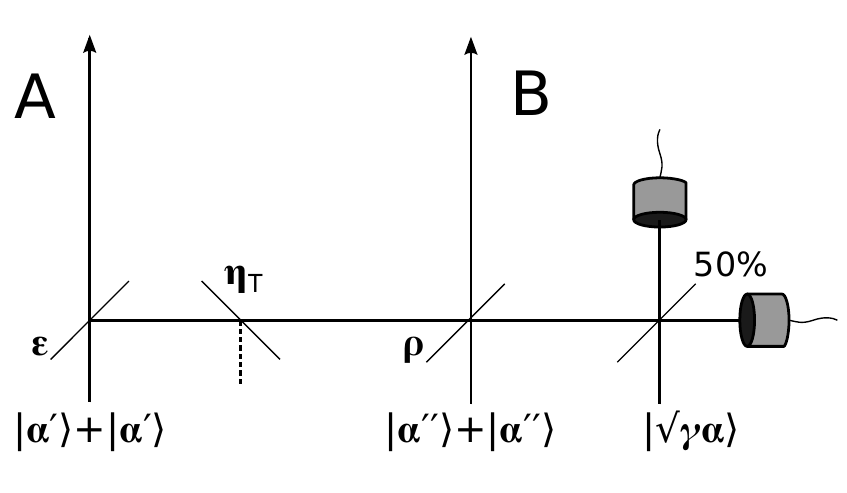}}
  \caption{Our proposed entanglement distribution scheme.  The parameters
	  $\alpha^\prime$ and $\eta_T$ are the same as before and defined in
	  Eqs.~(\ref{aprime}) and (\ref{etat}).  New parameters
	  $\alpha^{\prime\prime}$, $\rho$ and $\gamma$ are defined in
	  Eqs.~(\ref{app}), (\ref{rho}) and (\ref{gammaeqn}).  Just as before
	  each detector has a loss of $l$ and the encoding of the heralded ECS
	  is $\alpha$.  This scheme only has one success signal and on
	  achieving it an ECS with a positive superposition and in phase
          coherent states is produced.} 
  \label{proposed_scheme}
\end{figure}

\begin{figure*}[ht]
  \centerline{\includegraphics[width=15cm]{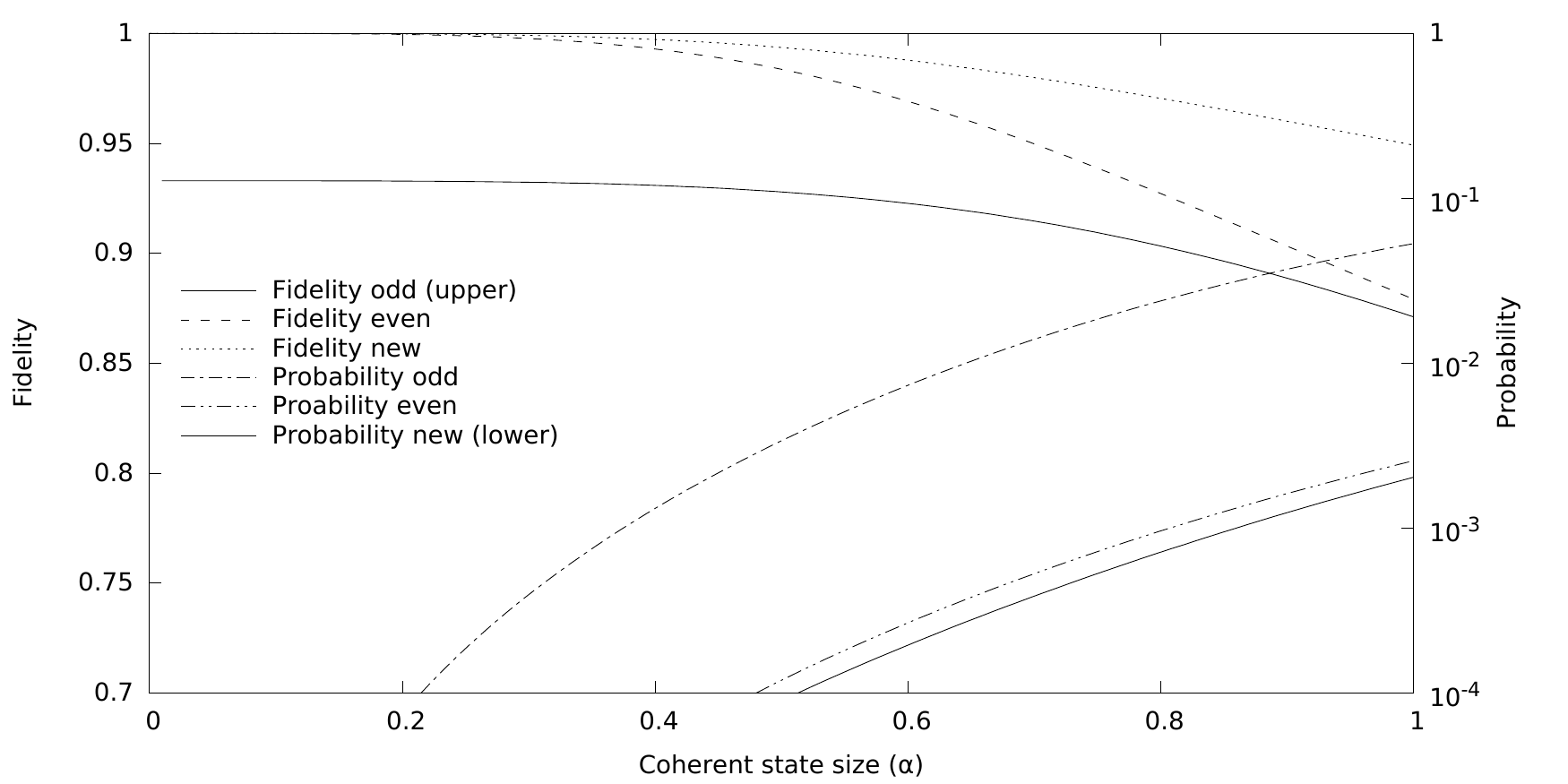}}
  \caption{Fidelity (top curves, LHS axis) and probability of success (bottom
	  curves, RHS axis) as a function of the output encoding coherent state
	  amplitude $\alpha$.  Parameters used for these plots are $\epsilon =
	  0.1$, $\eta_T = 0.5$, $l=0.5$. `Odd' and `even' refer to the scheme
	  from~\cite{grangier:nature} when the post-selection is done by
	  counting an odd number of photons or an even but non-zero number of
	  photons respectively. `New' refers to the new scheme presented here.}
  \label{apl_graph}
\end{figure*}

The fidelity of this new scheme is given by
\begin{equation}
  \mathcal{F}_{new}(\alpha) = 
  \frac{1}{(1+\tanh(\epsilon\eta_T\alpha^2)\tanh(2(1-\epsilon)\alpha^2))}
\label{eq:fnew}
\end{equation}
and the probability of success
\begin{equation}
  P_{new}(\alpha) = 
    \frac{(1-\exp(-(1-l)\gamma^2/2))^2}
         {2(1+\cosh((1-\rho)\alpha^2)/\cosh((1+\rho)\alpha^2))}.
  \label{eq:pnew}
\end{equation}
The details of this calculation are similar to that which were performed for
the other scheme and are contained in Appendix~\ref{appendix2}.
Figure~\ref{apl_graph} compares these new values for fidelity and probability
against those computed for the previous scheme using the same channel and
detector parameters.  This plot shows that for a fixed channel loss the
fidelity of achieving the desired coherent state size is higher for all
amplitudes when using the new scheme.  However, the probability is lower than
the probabilities of success for the old scheme.

\begin{figure}[hb]
  \centerline{\includegraphics{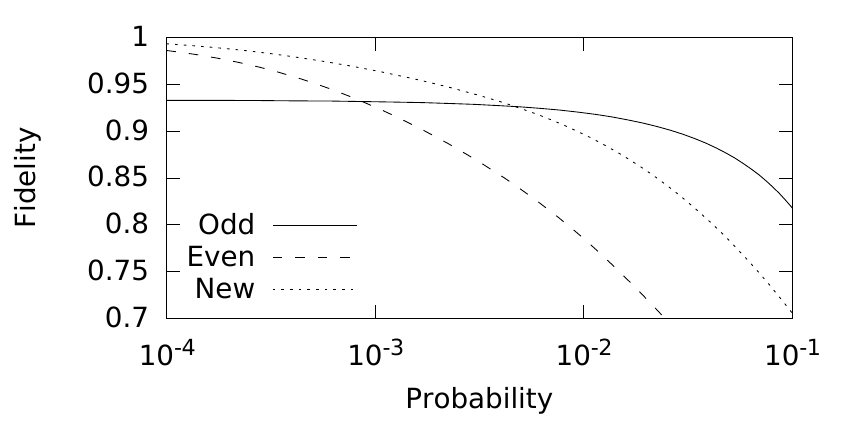}}
  \caption{Parametric plot of fidelity and probability using coherent state
	  amplitude as the parameter.  This plot was generated for $\epsilon =
	  0.1$, $\eta = 0.5$ and $l=0.5$ as in Fig.~\ref{apl_graph}.  For the
	  low probability events, which correspond to smaller choices of
	  $\alpha$ we see that the new scheme has a region in which it is
	  better in both probability and fidelity for this choice of
	  parameters.  For the choice of parameters in this plot, the region of
	  higher probability and fidelity of our new scheme is for $\alpha \lessapprox 1.2$.}
  \label{para_graph}
\end{figure}

If the encoding amplitude of the output entangled state is not required to be a
particular value, then this suggests a trade off would be possible between
fidelity and probability.  To make this comparison we plot in
Fig.~\ref{para_graph} a parametric plot of fidelity and probability where the
parameter defining the curve is $\alpha$.  Under this relaxed comparison we can
see that there is a region when the encoding size is small in which for a given
probability of success the fidelity of the new scheme is superior.

\section{Conclusion}

We have proposed and analysed a scheme for generating entangled coherent states
over a lossy environment which utilises a loss tolerant unambiguous detection
scheme to perform the heralding measurement.  We have compared this scheme to
one designed for a similar purpose in Ref.~\cite{grangier:nature} which uses
photon number resolving detectors  for heralding. We find that for any given
coherent state encoding amplitude, our scheme gives a better fidelity, but a
lower probability of success.  However, if comparison is made where the
encoding amplitude is not required to be the same, we find that for a given
probability of success our scheme can give a higher fidelity.  In particular,
when channel loss and detector loss are both $50\%$, a better fidelity for the
same probability is possible with our new scheme when preparing an ECS with
$\alpha \lessapprox 1.2$.  Our study may be useful for long-distance quantum
communication using SCSs and realistic detectors.

\acknowledgements This work was supported by the Australian Research Council
Centre of Excellence for Quantum Computation and Communication Technology
(Project number CE110001027) and by the National Research Foundation of Korea
(NRF) grant funded by the Korea government (MSIP) (No. 2010-0018295).

\begin{widetext}

\appendix
\section{Analysis of Ourjoumtsev \textbf{\textit{et al}}.'s Scheme}
\label{appendix:original}

We here present a detailed analysis of the scheme in
Ref.~\cite{grangier:nature} to find the fidelity and the success probability.
The methodology for the calculation is quite simple, however the equations that
result can be long and difficult to work with.  There are three steps to the
process.  First SCSs are prepared.  Next the prepared states are evolved in a
linear optical network.  Finally, some of the output modes are detected.  The
calculation we wish to perform is to find the output state from the modes which
are not detected conditional on the particular results from the measured modes.

The output state, including the amplitude and phase terms, for each coherent
states which make up the basis states in the coherent state superposition are
calculated.  Then to calculate the output state for the input coherent state
superposition, the terms are added together.  The probability can then be
computed as the normalisation of this resultant state.  The fidelity can then
be calculated by computing the overlap between renormalised state and the
desired entangled coherent state superposition.

The evolution of coherent states through a linear optical network is exactly
that which would result from the classical field amplitudes with the coherent
state eigenvalue playing the role of the field amplitude.  The evolution on the
coherent state basis states for the scheme from~\cite{grangier:nature} can be
calculated from the schematic shown in Fig.~\ref{grangier_experiment}.  This
linear network with loss $\eta$ on both halfs of the channel and detector loss
$l$, induces the transformation on the coherent states as
\[
\ket{\pm \sqrt{1+\ep}\alpha, \pm \sqrt{1+\ep}\alpha}
\rightarrow
\ket{
\pm\alpha,
\pm\alpha,
\pm\sqrt{2\ep(1-\eta)(1-l)}\alpha,
0,
\underline{\pm\sqrt{\ep\eta}\alpha},
\underline{\pm\sqrt{\ep\eta}\alpha},
\underline{\pm\sqrt{2\ep(1-\eta)l}\alpha},
\underline{0}
},
\]
\[
\ket{\pm\sqrt{1+\ep}\alpha, \mp\sqrt{1+\ep}\alpha}
\rightarrow
\ket{
\pm\alpha,
\mp\alpha,
0,
\pm\sqrt{2\ep(1-\eta)(1-l)}\alpha,
\underline{\pm\sqrt{\ep\eta}\alpha},
\underline{\mp\sqrt{\ep\eta}\alpha},
\underline{0},
\underline{\pm\sqrt{2\ep(1-\eta)l}\alpha}
}.
\]
where
\[
\ep = \frac{\epsilon}{1-\epsilon}
\]
and $\epsilon$ is the initial tapping off ratio.  On the right hand side of the
equation we have chosen the ordering of modes to be: A output, B output, C
output 1 (detected), C output 2 (detected), A channel loss mode, B channel
loss mode, C detector loss 1, C detector loss 2.  Detector modes will be
projected onto the Fock basis state corresponding to the detection results and
the loss modes, shown underlined, will be traced out.

C detects $n$ and $m$ in the detector outputs and $a$,$b$,$c$,$d$ photons are
present in the environment (which will eventually be summed over to perform the
trace), then we have the transformation including the amplitude and phase
factors
\[
\ket{\sqrt{1+\ep}\alpha,\sqrt{1+\ep}\alpha}\rightarrow
e^{-\ep\alpha^2}
\frac{\left(\sqrt{2\ep(1-\eta)(1-l)}\alpha\right)^n}{\sqrt{n!}} 
\delta_{m,0}
\frac{\left(\sqrt{\ep\eta}\alpha\right)^{a+b}}{\sqrt{a!b!}}
\frac{\left(\sqrt{2(1-\eta)\ep l}\alpha\right)^c}{\sqrt{c!}}
\delta_{d,0}
\ket{\alpha, \alpha},
\]
\[
\ket{-\sqrt{1+\ep}\alpha,-\sqrt{1+\ep}\alpha}\rightarrow
e^{-\ep\alpha^2}
\frac{\left(\sqrt{2\ep(1-\eta)(1-l)}\alpha\right)^n}{\sqrt{n!}} (-1)^n
\delta_{m,0}
\frac{\left(\sqrt{\ep\eta}\alpha\right)^{a+b}}{\sqrt{a!b!}} (-1)^{a+b}
\frac{\left(\sqrt{2\ep(1-\eta)l}\alpha\right)^c}{\sqrt{c!}} (-1)^c
\delta_{d,0}
\ket{-\alpha, -\alpha},
\]
\[
\ket{\sqrt{1+\ep}\alpha,-\sqrt{1+\ep}\alpha}\rightarrow
e^{-\ep\alpha^2}
\delta_{n,0}
\frac{\left(\sqrt{2\ep(1-\eta)(1-l)}\alpha\right)^m}{\sqrt{m!}} 
\frac{\left(\sqrt{\ep\eta}\alpha\right)^{a+b}}{\sqrt{a!b!}} (-1)^b
\delta_{c,0}
\frac{\left(\sqrt{2\ep(1-\eta)l}\alpha\right)^d}{\sqrt{d!}}
\ket{\alpha, -\alpha},
\]
\[
\ket{-\sqrt{1+\ep}\alpha,\sqrt{1+\ep}\alpha}\rightarrow
e^{-\ep\alpha^2}
\delta_{n,0}
\frac{\left(\sqrt{2\ep(1-\eta)(1-l)}\alpha\right)^m}{\sqrt{m!}} (-1)^m
\frac{\left(\sqrt{\ep\eta}\alpha\right)^{a+b}}{\sqrt{a!b!}} (-1)^a
\delta_{c,0}
\frac{\left(\sqrt{2\ep(1-\eta)l}\alpha\right)^d}{\sqrt{d!}} (-1)^d
\ket{\alpha, -\alpha},
\]
where $\delta_{x,y}$ is the Kronecker delta.  Now we can use linearity of the
evolution of the quantum state to compute the output given an input which is a
superposition of these coherent state inputs.  The normalized input state is
\begin{multline}
N_+(\sqrt{1+\ep}\alpha)^2
\left(\ket{\sqrt{1+\ep}\alpha}+\ket{-\sqrt{1+\ep}\alpha}\right) 
\otimes \left(\ket{\sqrt{1+\ep}\alpha} + \ket{-\sqrt{1+\ep}\alpha}\right)
= \\ N_+(\sqrt{1+\ep}\alpha)^2 \left(
\ket{\sqrt{1+\ep}\alpha,\sqrt{1+\ep}\alpha}
+\ket{-\sqrt{1+\ep}\alpha,\sqrt{1+\ep}\alpha}
+\ket{\sqrt{1+\ep}\alpha,-\sqrt{1+\ep}\alpha}
+\ket{-\sqrt{1+\ep}\alpha,-\sqrt{1+\ep}\alpha}
 \right)
\end{multline}
where
\[
N_\pm(\beta) = \frac{1}{\sqrt{2(1\pm e^{-2\beta^2})}}
\]
to which we then need to apply the above transformations.  We can see that
this results in the superposition of all of the above transformations.

This would result in a very long expression for the output state.  However, we
can look at particular values of the heralding signal to pick apart the
components.  If we consider the case where $m=0$ and $n\neq 0$ then the output
which results from the input components
$\ket{\sqrt{1+\ep}\alpha,-\sqrt{1+\ep}\alpha}$ and
$\ket{-\sqrt{1+\ep}\alpha,\sqrt{1+\ep}\alpha}$ have zero contribution to the
output state.  This is due to the coherent state amplitude at the detector
corresponding to the $n$ mode being zero.  Therefore for this case we only need
consider the top two transformations.  If we consider the case where $n$ is odd
and consider the traced out modes to be measured in the Fock basis (this choice
is arbitrary, as the trace is basis independent, but convenient) then we find
the output state is then 
\[ 
	N_+(\sqrt{1-\ep}\alpha)^2 e^{-\ep\alpha^2}
	\frac{\left(\alpha\sqrt{2\ep(1-\eta)(1-l)}\right)^n}{\sqrt{n!}}
	\frac{\left(\alpha\sqrt{\ep\eta}\right)^{a+b}}{\sqrt{a!b!}}
\frac{\left(\alpha\sqrt{2\ep(1-\eta)l}\right)^c}{\sqrt{c!}} \delta_{d,0}
\ket{(-1)^{a+b+c+1}\ _{\alpha}} 
\] 
where 
\[ 
	\ket{\pm_\beta} = \ket{\beta,\beta} \pm \ket{-\beta,-\beta} 
\] 
and is unnormalised.  The sign of the superposition is determined by the
numbers $a$,$b$ and $c$ which relate to the number of photons in the loss
modes.  We need to mix over these possibilities to complete the partial trace.
Each sum can be computed individually.  Take for example the sum over $c$.
There are two cases, where $c$ is odd and where $c$ is even (including zero).
When it is even, the $\ket{-\alpha,-\alpha}$ state contributes no minus sign
relative to the $\ket{\alpha,\alpha}$ state and the output state is the same
for all even cases, though the amplitude is different.  When it is odd, there
is a sign change, but again the state is the same for all odd cases.  When the
sum is performed the coefficients sum to a hyperbolic function.  This same
effect will occur for the $a$ and $b$ summations as well.  The sign changes can
be combined as two minus signs cancel.

Applying these sums and the hyperbolic trigonometric double angle formula,
results in the output state
\begin{equation}
  N_+(\sqrt{1+\ep}\alpha)^4 
  e^{-2\ep\alpha^2} 
  \frac{(2\ep(1-\eta^\prime)\alpha^2)^n}{n!}
  \left(
    \sinh(2\ep\eta^\prime\alpha^2) 
    \ket{+_\alpha}\bra{+_\alpha} + \cosh(2\ep\eta^\prime\alpha^2)
    \ket{-_\alpha}\bra{-_\alpha}
  \right) \nonumber
\end{equation}
where
\begin{equation}
  \eta^\prime = 1-(1-l)\sqrt{1-\eta_T}
\end{equation}
where $\eta_T$ is the total loss between the two parties.  The number
$\eta^\prime$ is the combination of half the channel loss and loss from one of
the detectors.  Note that when $m$ is odd and $n=0$ we obtain the same formula
for the even case if one of the modes is phase shifted by $\pi$ as merely the
other two components of the input state are selected out.  This results in
effectively multiplying the even and odd case density operators by two.
Finally, for the even case that the result will be an expression of the same
form but with the minus and plus states swapped.  The only case in which an
expression like this is not applicable is when $m=0$ and $n=0$.  We can now
sum over those states where the output is identical and assume that the easy
phase shift correction has been applied (but not to the superposition sign
change) to result in two expressions, one for the case of an even (but
non-zero) result
\begin{equation}
\rho_{even} = N_+(\sqrt{1+\ep}\alpha)^4 
e^{-2\ep\alpha^2} (\cosh (2\ep(1-\eta^\prime)\alpha^2)-1)
\left(
\cosh(2\ep\eta^\prime\alpha^2) \ket{+_\alpha}\bra{+_\alpha} 
+ \sinh(2\ep\eta^\prime\alpha^2) \ket{-_\alpha}\bra{-_\alpha}
\right)
\end{equation}
and for the odd result
\begin{equation}
\rho_{odd} = N_+(\sqrt{1+\ep}\alpha)^4 
e^{-2\ep\alpha^2} \sinh (2\ep(1-\eta^\prime)\alpha^2) \left(
\sinh(2\ep\eta^\prime\alpha^2)\ket{+_\alpha}\bra{+_\alpha}
+\cosh(2\ep\eta^\prime\alpha^2)\ket{-_\alpha}\bra{-_\alpha} \right).
\end{equation}
It is important to note that in the notation used here the plus and the minus
kets are unnormalised entangled states.  To calculate probability and fidelity
it is important to keep this in mind or factor out a normalisation coefficient
which will in general not be $1/\sqrt{2}$ do to the overlap of the states
forming the superposition.  So from these expressions we obtain the
probabilities of success in Eqs.~(\ref{eq:pe}) and (\ref{eq:po}) from the trace
% factor of two from two possiblilites
%\begin{equation}
%	P_{even} = \frac{e^{-2 \ep \alpha^2}}{2(1+e^{-2 (1+\ep) \alpha^2})}%
%	\left( \cosh(2 \ep (1 - \eta^\prime)\alpha^2) - 1 \right)
%	\left(
%	2 \cosh(2 \ep \eta^\prime \alpha^2) (1+e^{-2\alpha^2}) +
%	2 \sinh(2 \ep \eta^\prime \alpha^2) (1-e^{-2\alpha^2})
%	\right)
%\end{equation}
%\begin{equation}
%	P_{odd} = \frac{e^{-2 \ep \alpha^2}}{2(1+e^{-2 (1+\ep) \alpha^2})}
%	\sinh(2 \ep (1 - \eta^\prime)\alpha^2) 
%	\left(
%	2 \sinh(2 \ep \eta^\prime \alpha^2) (1+e^{-2\alpha^2}) +
%	2 \cosh(2 \ep \eta^\prime \alpha^2) (1-e^{-2\alpha^2})
%	\right)
%\end{equation}
and the fidelities in Eqs.~(\ref{eq:fe}) and (\ref{eq:fo}) are calculated by using the probabilities of success to
renormalise the density operator, then take the appropriate coefficients (the
minus superposition for odd and plus superposition for even).
% to arrive at the
%fidelity
%\begin{equation}
%\mathcal{F}_{even} = (1+\tanh(2\ep\eta^\prime\alpha^2)\tanh(2\alpha^2))^{-1}
%\end{equation}
%\begin{equation}
%\mathcal{F}_{odd} = (1+\tanh(2\ep\eta^\prime\alpha^2)\coth(2\alpha^2))^{-1}.
%\end{equation}

%\appendix
\section{Proposed Scheme}
\label{appendix2}

The calculation of the fidelity and probability for the scheme proposed in
this paper proceeds in much the same way as that given above.  First we write
down the transformation of the basis coherent states through the linear
network including the lossy channel.
\[
\ket{\pm \sqrt{1+\ep} \alpha, \pm\sqrt{1-\rho}\alpha, \sqrt{\gamma}\alpha}
\rightarrow
\ket{\pm \alpha, \pm \alpha, \underline{\pm \sqrt{\ep\eta_T} \alpha}, 
  \sqrt{\frac{\gamma}{2}}\alpha,  \sqrt{\frac{\gamma}{2}}\alpha}
\]
\[
\ket{\pm \sqrt{1+\ep} \alpha, \mp\sqrt{1-\rho}\alpha, \sqrt{\gamma}\alpha}
\rightarrow
\ket{\pm \alpha, \pm 2\rho\alpha, \underline{\pm \sqrt{\ep\eta_T} \alpha}, 
  \begin{array}{cc}
    0 \\ \sqrt{2}\gamma
  \end{array},
  \begin{array}{cc}
     \sqrt{2} \gamma \\ 0
  \end{array}
}
\]
where
\[
\rho = \ep (1-\eta_T)
\]
which is also the reflectivity of the beamsplitter for side B and
\[
\gamma = 4 \rho(1-\rho).
\]
The mode ordering on the right hand side is: A output, B output, channel loss,
B detector 1, B detector 2.  The channel loss mode is underlined as it is to be
traced out. The mode ordering on the left hand side is: A input, B input and B
detector input.  There are input modes associated with the loss mode and with
the beam-splitter with A which are always prepared in the vacuum state and are
not shown.  In the case where the signs of the input coherent states are the
same, perfect interference occurs before the unambiguous state discrimination
scheme and results in the vacuum being input, hence the coherent state splits
evenly and with the same sign.   This occurs by design and is why the
parameters take the values given above.   In the other two cases the input is
either the plus or minus $\gamma$ coherent state and hence the usual situation
for the unambiguous state discrimination applies.

These transformations are then applied to the input SCS states
\begin{multline}
N_+(\sqrt{1+\ep}\alpha) N_+(\sqrt{1-\rho}\alpha)
(\ket{\sqrt{1+\ep}\alpha} + \ket{-\sqrt{1+\ep}\alpha})\otimes
(\ket{\sqrt{1-\rho}\alpha} + \ket{\sqrt{1-\rho}\alpha}) = \\
N_+(\sqrt{1+\ep}\alpha) N_+(\sqrt{1-\rho}\alpha)
\left(
\ket{\sqrt{1+\ep}\alpha,\sqrt{1-\rho}\alpha}
+\ket{\sqrt{1+\ep}\alpha,-\sqrt{1-\rho}\alpha}\right.\\ \left.
+\ket{-\sqrt{1+\ep}\alpha,\sqrt{1-\rho}\alpha}
+\ket{-\sqrt{1+\ep}\alpha,-\sqrt{1-\rho}\alpha} 
\right).
\end{multline}
Successful detection occurs when both of the detectors record a click of any
number of photons.  This will clearly select out the cases where the sign of
the coherent state in both A and B's output is the same.  Proceeding as before
we will compute the partial trace in the Fock basis.  So, if $a$ photons are in
the loss mode and the detectors register $n\neq0$ and $m\neq0$ photons the
output state is
\[
N_+(\sqrt{1+\ep}\alpha)N_+(\sqrt{1-\rho}\alpha)
e^{-(\eta_T\ep + \gamma)\alpha^2/2} 
\frac{1}{\sqrt{n!m!}}\left(\sqrt{\frac{\gamma}{2}}\alpha\right)^{n+m}
\frac{(\sqrt{\eta_T\ep}\alpha)^a}{\sqrt{a!}}
\ket{{(-1)^a}_\alpha}.
\]
The reduced density matrix when tracing out $a$ is
\[
N_+(\sqrt{1+\ep}\alpha)^2 N_+(\sqrt{1-\rho}\alpha)^2
(1-e^{-\gamma\alpha^2/2})^2
e^{-\eta_T\ep\alpha^2}
\left(
\cosh(\eta_T\ep\alpha^2)\ket{+_{\alpha}}\bra{+_{\alpha}} +
\sinh(\eta_T\ep\alpha^2)\ket{-_{\alpha}}\bra{-_{\alpha}} 
\right)
\]
%\[
%P_{new} = \frac{(1-e^{-\gamma\alpha^2/2})^2}{2} \left(
%1+\frac{\cosh(\ep(2-\eta_T)\alpha^2)}{\cosh((2+\ep\eta_T)\alpha^2)}
%\right)^{-1}
%\]
%\[
%\mathcal{F}_{new} = \left(1+\tanh(\ep\eta_T\alpha^2)\tanh(2\alpha^2)\right)^{-1}
%\]
that we obtain the success probability and fidelity in Eqs.~(\ref{eq:fnew}) and
(\ref{eq:pnew}).  Incorporating detector loss for the success case is simply a
matter of reducing the coherent state amplitude by a factor of $\sqrt{1-l}$ as
the state incident upon the detector is a coherent state and loss will merely
reduce the coherent state amplitude and no superpositions of coherent states
occur when a successful signal is achieved.  This can be achieved by adjusting
$\gamma$ so that it is $\sqrt{1-l}$ times smaller.  This then recreates
equation~\ref{gammaeqn}.  Here we can see that the fidelity is independent of
$\gamma$, one of the critical features of our design.

% Fidelity_new(alpha) = (1+tanh(eps*eta*alpha**2)*tanh(2*(1-eps)*alpha**2))**(-1)
% Probability_new(alpha) = (1-exp(-(1-l)*gamma**2*alpha**2/2))**2/2/(1+cosh((1-rho)*alpha**2)/cosh((1+rho)*alpha**2))

\end{widetext}

\end{document}